\documentclass[12pt]{article}
\usepackage{axodraw,bbold}

\catcode`@=12
\begin{document}
\thispagestyle{empty}
\begin{flushright} 
UCRHEP-T433\\ 
July 2007\
\end{flushright}
\vspace{0.5in}
\begin{center}
{\LARGE	\bf Gluino Axion, Neutrino Seesaw,\\ 
Dynamical Gaugino Mass,\\
and $A \simeq 0$ Supersymmetry\\}
\vspace{1.5in}
{\bf Ernest Ma\\}
\vspace{0.2in}
{\sl Department of Physics and Astronomy, University of 
California,\\ Riverside, California 92521, USA\\}
\vspace{1.0in}
\end{center}

\begin{abstract}\
In the axionic solution of the strong CP problem, fermions which transform 
under quantum chromodynamics (QCD) are required.  In supersymmetry, by 
equating $U(1)_{PQ}$ with $U(1)_R$, the natural candidates are the gluinos, 
as pointed out some years ago.  A new specific implementation of this idea 
is proposed, linking the gluino axion scale to that of the canonical seesaw 
mechanism for neutrinos.  Gaugino masses are generated dynamically and the 
$A$ term is predicted to be very small.
\end{abstract}

\newpage
\baselineskip 24pt

The axion is a nearly massless pseudoscalar particle postulated to solve 
the strong CP problem \cite{p06}.  As such, it must be related to the 
mass-generation mechanism of a colored fermion multiplet.  Instead of 
quarks, it was first pointed out by Demir and Ma \cite{dm00,dm01} that gluinos 
may also be used.  In this paper, a new specific implementation of this idea 
is proposed, where the gluino axion scale and that of the canonical seesaw 
mechanism for neutrinos in supersymmetry are one and the same 
\cite{dms00,m01,m03}. As a consequence, gaugino masses and the $A$ term 
in supersymmetry are forbidden at tree level.  New singlet heavy quarks 
(such as those available in the \underline{27} representation of $E_6$) 
are introduced to allow the gluino mass to be generated in one loop.  
The $A$ term is also radiatively generated but remains negligible.

The axion to be discussed is a singlet under the $SU(3)_C \times SU(2)_L 
\times U(1)_Y$ gauge group of the Standard Model (SM).  It comes from the 
spontaneous breaking of an anomalous global symmetry, i.e. $U(1)_{PQ}$, the 
choice of which defines the model.  If it is identified \cite{dm00,dm01,dms00} 
with the $U(1)_R$ of supersymmetric transformations, then the resulting 
axion couples to gluinos, not quarks. Under $U(1)_R$, the scalar components 
of a chiral superfield transform as $\phi \to e^{i \theta R} \phi$, 
whereas the fermionic components transform as $\psi \to e^{i \theta (R-1)} 
\psi$.  For the Lagrangian to be invariant under $U(1)_R$, the superpotential 
$\hat W$ should have $R=2$.  In the Minimal Supersymmetric Standard Model 
(MSSM), this is explicitly broken by the term $\mu \hat H_u \hat H_d$, 
resulting in the conservation of only its well-known discrete remnant, 
i.e. $R$ parity.

The first task is to devise a mechanism for having the axion scale \cite{r06} 
at $10^{11}$ GeV or so and yet for it to be related to the gluino mass at 
the electroweak scale.  Following Ref.~\cite{dm01}, consider three singlet 
superfields $\hat S_2$, $\hat S_1$, $\hat S_0$ with $U(1)_R$ charges 2, 1, 0 
and transforming under an additonal discrete $Z_3$ symmetry as $\omega^2$, 
$\omega$, $\omega$ where $\omega = \exp(2 \pi i/3)$. The most general 
$R=2$ superpotential is given by
\begin{equation}
\hat W = m_2 \hat S_2 \hat S_0 + f_1 \hat S_1 \hat S_1 \hat S_0 + 
\Lambda \hat S_2 + m_1 \hat S_1 \hat S_1,
\end{equation}
where $Z_3$ is broken only by the \underline{soft} terms $\Lambda \hat S_2$ 
and $m_1 \hat S_1 \hat S_1$.  The resulting scalar potential
\begin{equation}
V = |m_2S_2+f_1S_1^2|^2+|2m_1S_1+2f_1S_1S_0|^2+|\Lambda+m_2S_0|^2
\end{equation}
has a minimum at $V=0$ if
\begin{eqnarray}
v_2 &=& -{f_1 v_1^2 \over m_2}, \\
v_0 &=& -{m_1 \over f_1} = -{\Lambda \over m_2},
\end{eqnarray}
where $v_{2,1,0}$ are the vacuum expectation values of $S_{2,1,0}$ 
respectively.  Therefore, if $\Lambda$ is set equal to $m_1 m_2/f_1$, $U(1)_R$ 
may be broken spontaneously without breaking the supersymmetry.  This is of 
course fine tuning, but once it is done, soft supersymmetry breaking terms 
at the TeV scale will not change the basic quadratic relationship between 
$v_1$ and $v_2$ in the above. This allows $v_2$ to be much smaller than 
$v_1$ and is also the key to equating the axion scale to the neutrino seesaw 
mass scale, as shown below.

Consider now the superfields of the MSSM. Under $U(1)_R \times Z_3$, the 
Higgs superfields $\hat H_u$, $\hat H_d$ transform as $(0,\omega^2)$; 
$\hat Q = (\hat u, \hat d)$, $\hat L = (\hat \nu, \hat e)$ as $(3/2,1)$; 
$\hat u^c$, $\hat d^c$, $\hat e^c$, $\hat N^c$ as $(1/2,\omega)$.  [This 
differs from the usual $U(1)_R$ assignment by the transformation $R \to R + 
(3B+L)/2.$]  The resulting $R=2$ superpotential is given by
\begin{eqnarray}
\hat W &=& h_u \hat H_u \hat Q \hat u^c + h_d \hat H_d \hat Q \hat d^c + 
h_e \hat H_d \hat L \hat e^c \nonumber \\ 
&+& h_2 \hat S_2 \hat H_u \hat H_d + h_N \hat H_u \hat L \hat N^c + 
{1 \over 2} h_1 \hat S_1 \hat N^c \hat N^c.
\end{eqnarray}
The usual $\mu$ term is now replaced by $h_2 v_2$ and the singlet neutrino 
mass $m_N$ by $h_1 v_1$.

Using Eqs.~(3) and (4), with the redefinition of $\hat S_{2,1,0} \to 
v_{2,1,0} + \hat S_{2,1,0}$, Eq.~(1) can be rewritten as
\begin{equation}
\hat W = {m_2 \over v_1} (v_1 \hat S_2 - 2 v_2 \hat S_1) \hat S_0 + 
f_1 \hat S_1 \hat S_1 \hat S_0,
\end{equation}
showing clearly that the linear combination
\begin{equation}
\hat S = {v_1^* \hat S_1 + 2 v_2^* \hat S_2 \over \sqrt{|v_1|^2+4|v_2|^2}}
\end{equation}
is a massless superfield.  Consider now the breaking of supersymmetry by 
soft terms at the TeV scale which preserve the $U(1)_R$ symmetry but not 
necessarily the $Z_3$ discrete symmetry.  In the scalar sector, the 
important terms are $\mu_1^2 S_1^* S_1 + \mu_{12} S_1^2 S_2^* + \mu_{12}^* 
(S_1^*)^2 S_2$, which lift the indeterminacy \cite{m99} of Eq.~(3) and 
result in \cite{dm01}
\begin{equation}
|v_1|^2 = {\mu_1^2 \over 4 Re(\mu_{12}^* f_1/m_2)}.
\end{equation}
For example, let $m_2 = 10^{16}$ GeV, $f_1 = 0.1$, $\mu_1 = 20$ TeV, 
$\mu_{12} = 1$ TeV, then
\begin{equation}
v_1 = 10^{11}~{\rm GeV}, ~~~ v_2 = 10^5~{\rm GeV}.
\end{equation}
The neutrino seesaw mass scale $m_N = h_1 v_1$ may then be easily of order 
$10^8$ GeV if $h_1 \sim 10^{-3}$.

Since $v_1 >> v_2$, the scalar component (saxion) of the axion superfield 
is mostly $S_1$ and acquires a mass given by
\begin{equation}
m_S^2 = \mu_1^2 - 2 Re(\mu_{01} m_1/f_1),
\end{equation}
where the second term comes from $(\mu_{01} S_0 + \mu_{01}^* S_0^*) S_1^* 
S_1$.  As for the fermionic component (axino), since $\tilde S_1 \tilde S_1$ 
is an allowed term under $U(1)_R$, it can have an arbitrary Majorana mass 
at the scale of soft supersymmetry breaking.  In Eq.~(5), since $N^c$ is 
heavy and $h_2 v_2 = \mu$, the only term beyond those of the MSSM is
\begin{equation}
{2 \mu \over |v_1|} \hat S \hat H_u \hat H_d.
\end{equation}
If the axino is light enough, the would-be lightest supersymmetric particle 
of the MSSM will decay into it, allowing for possible collider signatures 
\cite{fhw05}.

The requirement of $U(1)_R$ symmetry forbids all gaugino masses at tree level 
as well as the trilinear scalar $A$ terms of the MSSM.  Whereas $A=0$ is not 
a problem phenomenologically, the absence of gaugino masses is not acceptable. 
Indeed, a mass for the gluino is necessary for the axion to couple to it. 
The direct coupling $S_2^* \tilde g \tilde g$ is not allowed because it is 
a hard term (of dimension four) which breaks supersymmetry.  Hence new 
singlet heavy quarks $h$, $h^c$ of charge $\mp 1/3$ are proposed, transforming 
under $U(1)_R \times Z_3$ as $(3/2,1)$, $(1/2,\omega)$ respectively. [They 
can come from the \underline{27} representaion of $E_6$ for example.] 
Since $d^c$ also transforms as $(1/2,\omega)$, there are two more terms 
in the superpotential, i.e.
\begin{equation}
m_h \hat h \hat h^c + h_{dh} \hat H_d \hat Q \hat h^c.
\end{equation}
The first term serves to define $\hat h^c$ and the second mixes $d$ 
and $h$, thus allowing $h$ to decay.  For $m_h$ large compared to 
the electroweak scale, this mixing is also small enough to be acceptable 
phenomenologically.  Now there can be a soft supersymmetry breaking 
trilinear scalar term $\lambda_h S_2^* \tilde h \tilde h^c$, which allows 
the gluino to acquire a mass in one loop as shown in Fig.~1.  The same 
mechanism also works for the $U(1)_Y$ gaugino.

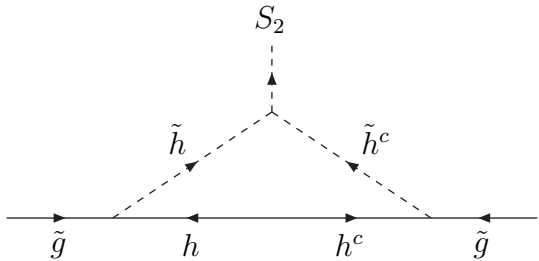
\begin{figure}[htb]
\begin{center}
\begin{picture}(360,120)(0,0)
\ArrowLine(80,10)(120,10)
\ArrowLine(180,10)(120,10)
\ArrowLine(180,10)(240,10)
\ArrowLine(280,10)(240,10)
\DashArrowLine(120,10)(180,50)3
\DashArrowLine(240,10)(180,50)3
\DashArrowLine(180,50)(180,75)3

\Text(100,0)[]{$\tilde g$}
\Text(260,0)[]{$\tilde g$}
\Text(150,0)[]{$h$}
\Text(210,0)[]{$h^c$}
\Text(145,40)[]{$\tilde h$}
\Text(220,40)[]{$\tilde h^c$}
\Text(180,85)[]{$S_2$}

\end{picture}
\end{center}
\caption{One-loop generation of gluino mass.}
\end{figure}

Assuming the mass eigenvalues of the $(\tilde h,\tilde h^c)$ sector to 
be $m_h^2 \pm |\lambda_h v_2^*|$, the gluino mass is given by
\begin{equation}
m_{\tilde g} = {\alpha_s m_h \over 16 \pi} \left[ {\ln(1-x^2) \over x} + 
\ln \left( {1+x \over 1-x} \right) \right],
\end{equation}
where $x = |\lambda_h v_2^*|/m_h^2$.  Let $m_h = 1.1 \times 10^5$ GeV, 
$\lambda_h = v_2 = 10^5$ GeV, $\alpha_s = 0.12$, then 
$m_{\tilde g} = 253$ GeV.

As for the Higgs sector, the important $B$ term is allowed by $U(1)_R$ 
symmetry, i.e. $\mu_{12}^2 H_u H_d + H.c.$  Together with the induced 
$\mu$ term, the $SU(2)_L$ and $U(1)_Y$ gauginos also receive radiative 
mass contributions as shown in Fig.~2.  Note that there are additional 
heavy Higgs superfields beyond those of the MSSM which are available for 
example in $E_6$, allowing these masses to be also of order $m_{\tilde g}$.  
Once the gauginos are massive, the $A$ term is also radiatively generated, 
but it is a two-loop effect, hence $A \simeq 0$ is expected.

\begin{figure}[htb]
\begin{center}
\begin{picture}(360,120)(0,0)
\ArrowLine(80,10)(120,10)
\DashArrowLine(120,10)(180,10)3
\DashArrowLine(240,10)(180,10)3
\ArrowLine(280,10)(240,10)
\ArrowLine(180,50)(120,10)
\ArrowLine(180,50)(240,10)
\DashArrowLine(180,50)(180,75)3

\Text(100,0)[]{$\tilde w$}
\Text(260,0)[]{$\tilde w$}
\Text(150,0)[]{$H_u$}
\Text(210,0)[]{$H_d$}
\Text(140,40)[]{$\tilde H_u$}
\Text(220,40)[]{$\tilde H_d$}
\Text(180,85)[]{$S_2$}

\end{picture}
\end{center}
\caption{One-loop generation of $SU(2)_L$ gaugino mass.}
\end{figure}
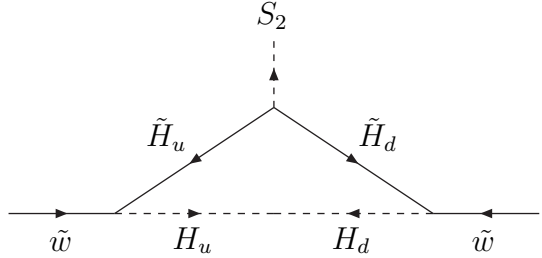

Because of the necessity of generating realistic gaugino masses, the 
scale of soft supersymmetry breaking as well as $v_2$ and the mass of new 
particles should be of order $10^5$ GeV. However, just as the SM allows 
a wide range of Yukawa couplings, some of the supersymmetry breaking  
parameters may be as low as $10^2$ GeV.  The experimental tests of this 
model are Eq.~(11) and the prediction $A \simeq 0$.  Without the $A$ term, 
the mixing matrix linking left sfermions with right sfermions is 
automatically proportional to the corresponding fermion mass matrix.  
This solves the usual problem of flavor changing neutral currents in 
supersymmetry.

I thank V. Barger, M. Frigerio, and K. Hagiwara for discussions during the 
2007 Neutrino Workshop at the Aspen Center for Physics. This work was 
supported in part by the U.~S.~Department of Energy under Grant 
No.~DE-FG03-94ER40837.

\newpage

\baselineskip 18pt

\bibliographystyle{unsrt}

\begin{thebibliography}{99}
\bibitem{p06} For a recent review, see for example R. D. Peccei, 
hep-ph/0607268.
\bibitem{dm00} D. A. Demir and E. Ma, Phys. Rev. {\bf D62}, 111901(R) (2000).
\bibitem{dm01} D. A. Demir and E. Ma, J. Phys. {\bf G27}, L87 (2001).
\bibitem{dms00} D. A. Demir, E. Ma, and U. Sarkar, J. Phys. {\bf G26}, L117 
(2000).
\bibitem{m01} E. Ma, Phys. Lett. {\bf B514}, 330 (2001).
\bibitem{m03} E. Ma, J. Phys. {\bf G29}, 313 (2003).
\bibitem{r06} See for example G. Raffelt, hep-ph/0611350.
\bibitem{m99} E. Ma, Mod. Phys. Lett. {\bf A14}, 1637 (1999).
\bibitem{fhw05} See for example B. Feldstein, L. J. Hall, and T. Watari, 
Phys. Lett. 
{\bf B607}, 155 (2005).
\end{thebibliography}

\end{document}